\documentclass[useAMS,usenatbib]{mn2e}

 \usepackage{times}

\usepackage{graphicx}
\usepackage{amsmath, amssymb}

\title[Rotational disruption and Main Belt asteroid sizes]{Effect of rotational disruption on the size-frequency distribution of the Main Belt asteroid population}
\author[Seth A. Jacobson, Francesco Marzari, Alessandro Rossi, Daniel J. Scheeres \& Donald R. Davis]{Seth A. Jacobson$^{1,2,3}$\thanks{E-mail:
seth.jacobson@oca.eu (SAJ)}, Francesco Marzari$^{4}$, Alessandro Rossi$^{5}$, Daniel J. Scheeres$^{6}$ \newauthor \& Donald R. Davis$^{7}$\\
$^{1}$Department of Astrophysical and Planetary Sciences, UCB 391, University of Colorado, Boulder, CO 80309-0391 USA\\
$^{2}$Laboratoire Lagrange, Observatoire de la C{\^o}te d'Azur, Boulevard de l'Observatoire, 06304 Nice Cedex 4, France\\
$^{3}$Bayerisches Geoinstitut, Universt{\"a}t Bayreuth, D-95440 Bayreuth, Germany\\
$^{4}$Dipartimento di Fisica, Universit{\`a} di Padova, 35131 Padova, Italy\\
$^{5}$IFAC-CNR, 50019 Sesto Fiorentino, Firenze, Italy\\
$^{6}$Department of Aerospace and Engineering Sciences, UCB 429, University of Colorado, Boulder, CO 80309-0429 USA\\
$^{7}$Planetary Science Institute, 1700 East Fort Lowell Road, Suite 106, Tucson, AZ 85719, USA}
\begin{document}

\date{}

\pagerange{\pageref{firstpage}--\pageref{lastpage}} \pubyear{2013}

\maketitle

\label{firstpage}

\begin{abstract}
The size distribution of small asteroids in the Main Belt is assumed to be determined by an equilibrium between the creation of new bodies out of the impact debris of larger asteroids and the destruction of small asteroids by collisions with smaller projectiles. However, for a diameter less than 6 km we find that YORP-induced rotational disruption significantly contributes to the erosion even exceeding the effects of collisional fragmentation. Including this additional grinding mechanism in a collision evolution model for the asteroid belt, we generate size-frequency distributions from either an accretional~\citep{Weidenschilling:2011kt} or an ``Asteroids were born big''~\citep{Morbidelli:2009dd} initial size-frequency distribution that are consistent with observations reported in~\citet{Gladman:2009cx}. Rotational disruption is a new mechanism that must be included in all future collisional evolution models of asteroids.
\end{abstract}


\section{Introduction}
\label{sec:introduction}
The size-frequency distribution of the Main Belt asteroid population is an equilibrium between destruction and creation. Destruction occurs through two mechanisms: collisions and rotational disruption. Both produce fragments--new asteroids of smaller sizes. The equilibrium established when considering only the role of collisions is well studied~\citep[e.g. ][]{BottkeJr:2005gd}, but the role of rotational disruption has yet to be explored. We use an asteroid rotational evolution model combined with a collision evolution model to produce a new size-frequency distribution that accounts for both mechanisms.
 
Rotational disruption is driven by the Yarkovsky-O'Keefe-Radzievskii-Paddack (YORP) effect~\citep{Rubincam:2000fg,Taylor:2007kp}, which can rotationally accelerate asteroids to their critical spin disruption rates~\citep{Bottke:2006en,Scheeres:2007io,Walsh:2008gk}. The YORP effect changes the spin rate:
\begin{equation}
\dot{\omega} = \frac{Y}{2 \pi \rho R^2} \left( \frac{F_\odot }{a_\odot^2 \sqrt{1 - e_\odot^2}} \right)
\label{eqn:yorp}
\end{equation}
where $Y$ is the YORP coefficient determined by the asymmetric shape of the asteroid, $\rho$ is the density, $R$ is the radius of the asteroid, $a_\odot$ and $e_\odot$ are the heliocentric semi-major axis and eccentricity, and $F_\odot = 10^{14}$ kg km s$^{-2}$ is the solar radiation constant~\citep{Scheeres:2007kv}. The YORP effect has a strong size dependence. If the YORP coefficient $Y > 0$, then the spin rate accelerates towards a critical surface disruption limit. 

\begin{figure}
\begin{center}
\includegraphics[width=\columnwidth]{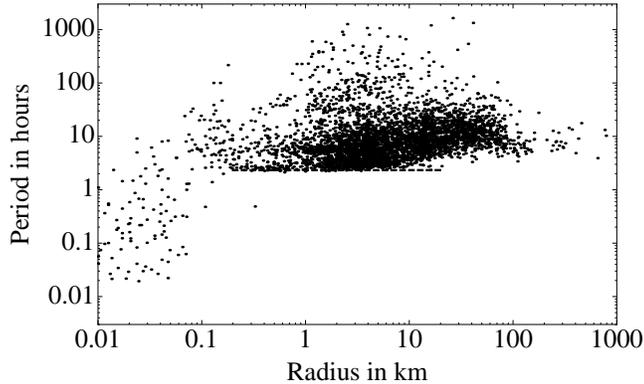}
\end{center}
\caption{Spin period distribution as a function of radius for near-Earth (NEA), Mars crossing (MCA) and Main Belt (MBA) asteroids as reported in the Asteroid Lightcurve Database~\citep{Warner:2009ds}. The dashed lines indicate the critical surface disruption period $P_d \sim 2.33$ h for radii $R > 250$ m.}
\label{fig:spindistribution}
\end{figure}
Rotational disruption occurs when centrifugal and gravitational accelerations become equal inside a rubble pile asteroid. Created by collisional processing, rubble pile asteroids are a collection of gravitationally bound boulders with a distribution of size scales and with very little or no tensile strength between them~\citep{Harris:1996tn,Asphaug:2002vp}. Evidence for rubble pile geophysics includes measured low bulk densities implying high porosities~\citep{Yeomans:1997fp,Ostro:2006dq}, the resolved surface of 243 Itokawa~\citep{Fujiwara:2006ca}, the observed critical spin limit amongst the asteroid population $P_d = \sqrt{3 \pi / \rho G} \sim$ 2.33 h (see Figure~\ref{fig:spindistribution}) where $G$ is the gravitational constant~\citep{Harris:1996tn,Pravec:2007ki}, and evidence that asteroid pairs form from rotational fission events~\citep{Pravec:2010kt}. Due to this strengthless internal structure, an asteroid eventually disrupts into components when it rotates at this disruption spin rate~\citep{Scheeres:2007io}. This simple story of rotational disruption is complicated by but reaffirmed when the asteroid's shape is allowed to evolve~\citep{Walsh:2008gk,Walsh:2012jt,Sanchez:2012hz,Sanchez:2013vm}. 

The disruption spin rate is also size dependent. Asteroids smaller than 250 m in radius are able to accelerate faster than the critical disruption period (see Figure~\ref{fig:spindistribution}). The strength holding these small bodies together is hypothesized to come from cohesive forces~\citep{Holsapple:2007eg,Scheeres:2012tj} if these bodies are still rubble piles or if these bodies are monolithic components, then the strength of the rock itself~\citep{Pravec:2000dr,Pravec:2007ki}. It is unclear what is happening at these small sizes.

Since the YORP effect is proportional to the radius squared (see Equation~\ref{eqn:yorp}), there is not a population of large asteroids spinning near the critical surface disruption limit (see Figure~\ref{fig:spindistribution}). We quantify this upper size limit by comparing rotational acceleration rates to collision rates using a timescale analysis. There are two possibly relevant collisional timescales: (1) the disruption timescale $\tau_\text{disr}$, how long before a collision occurs that removes more than half the mass of the asteroid, and (2) the rotational timescale $\tau_\text{rot}$, how long before a collision occurs that adds or subtracts angular momentum on the same order of magnitude as the asteroid's spin state.~\citet{Farinella:1998ff} provides an estimate of both timescales:
\begin{equation}
\tau_\text{disr} = 633\text{ My} \left( \frac{R}{1\text{ km}} \right)^{\frac{1}{2}}  \quad \tau_\text{rot} = 188\text{ My} \left( \frac{R}{1\text{ km}} \right)^{\frac{3}{4}} 
\end{equation}
We compare these timescales to an asteroid with a heliocentric orbit at $a_\odot = 2.5$ AU and a YORP coefficient of $Y = 0.01$:
\begin{equation}
\tau_{YORP} \sim \frac{ 2 \pi \omega_d \rho  R^2 a_\odot^2 }{Y F_\odot} =  42 \text{ My} \left( \frac{R}{1\text{ km}} \right)^2
\end{equation}
where $\omega_d = \sqrt{4 \pi \rho G / 3}$ is the critical disruption spin rate for a gravitationally bound spherical object. The timescale for YORP-induced rotational acceleration is always shorter than the collision-driven rotation timescale, and it is shorter than the collision-driven disruption timescale for asteroids with radii $R \lesssim$ 6 km. The YORP-induced rotational disruption timescale is longer than the age of the Solar System for $R \gtrsim$ 10 km explaining the lack of rapid rotators in Figure~\ref{fig:spindistribution} at large sizes. Therefore, we focus our rotational evolution model on asteroids with radii between 250 m and 15 km since in this range rotational disruption has a significant effect on the creation-destruction equilibrium which sets the size-frequency distribution.

\section{Methods}
\label{sec:asteroidpopulationevolutionmodel}
To understand the effects of rotational disruption on the evolution of the size-frequency distribution of the Main Belt, we have used two separate codes. The first model computes the frequency with which small ($0.25 < R < 15$ km) asteroids spin up to disruption. The second is a collisional evolution model where we have added the effects of rotational disruption by exploiting the outcome of the first code.

\subsection{The rotational evolution model}
The model computing the rotational evolution of small asteroids is a continuation of the code presented in~\citet{Marzari:2011dx}, which studied the rotational evolution of the Main Belt asteroid (MBA) population including both the YORP effect and collisions since both evolve the spin rate and direction. This was already an improvement and continuation of earlier projects by~\citet{Rossi:2009kz} and~\citet{Scheeres:2004bd}, which studied the near-Earth asteroid population. Similar to~\citet{Marzari:2011dx}, we use a Monte Carlo approach to individually simulate $2 \times 10^6$ asteroids for $4 \times 10^9$ years; this evolution assumes conditions that were only present after the late giant planet instability~\citep{Tera:1973tf,Levison:2011gt}. The spin rate and obliquity of each asteroid evolves constantly due to the YORP effect and collisions as in~\citet{Marzari:2011dx}. Unlike in the previous works, when the rotation rate of an asteroid exceeds a specified spin limit, the asteroid rotationally disrupts.

Since the exact rotational break-up spin rate is a complex function of the internal component distribution, the asteroid rotation evolution model utilizes the simple approximation that all ``rubble piles'' rotationally disrupt at the critical surface disruption spin limit for a ellipsoidal object: $\omega_d = S \sqrt{4 \pi \rho G/3}$ where $S$ is a shape factor determined by elliptic integrals from the semi-axes of the ellipsoidal figure~\citep{Scheeres:1994cb}. This approximation requires the system to rotationally accelerate for a longer period of time before undergoing rotational fission. With respect to the YORP timescale for rotational fission, this may accurately reflect delays in rotational fission due to shape evolution. Each asteroid is assigned also a shape from an ellipsoidal semi-axis ratio distribution for the purpose of calculating the critical spin limit. From largest to smallest, the tri-axial semi-axes are $a$, $b$, and $c$ and the axis ratios are drawn from normal distributions such that for $b/a$, the mean $\mu = 0.6$ with a standard deviation $\sigma = 0.18$ and for $c/a$, $\mu = 0.4$ and $\sigma = 0.05$~\citep{Giblin:1998io}. This shape distribution is in agreement with Hayabusa observations of boulders on 243 Itokawa and photometry of small, fast-rotating asteroids~\citep{Michikami:2010cr}, as well as agreement between the $b/a$ ratio and the mean amplitude of asteroids with diameters between 0.2 and 10 km~\citep{Pravec:2000dr}.

In addition to its shape, each asteroid is characterized by a number of fixed and evolving parameters including a diameter and a fixed semi-major axis $a_\odot$ and eccentricity $e_\odot$ from a Main Belt asteroid orbital element distribution. Both the YORP effect and collisions evolve the spin rate $\omega$ and the obliquity $\epsilon$ of each asteroid. The initial spin rate is drawn from a Maxwellian distribution with a $\sigma = 1.99$ corresponding to a mean period of $7.56$ h~\citep{Fulchignoni:1995um,Donnison:1999iv}.~\citet{Rossi:2009kz} demonstrated for models similar to the asteroid rotational evolution model that the steady-state spin rate distribution is independent of the initial spin rate distribution. We draw the initial obliquity of each asteroid from a flat distribution. The relative change in obliquity is used by the model to update the YORP coefficient, however the absolute obliquity is not currently used by the model. Thus the rotational evolution output is insensitive to the initial obliquity distribution, but it is a feature of the model that could be utilized in the future to compare input and output obliquity distributions.

In order to calculate the rotation evolution due to the YORP effect, each object is also assigned a non-dimensional YORP coefficient\footnote{\citet{Rossi:2009kz} and~\citet{Marzari:2011dx} notated the non-dimensional coefficient $Y$ as $C_Y$.} $Y$ from a gaussian distribution with a mean of $0$ and a standard deviation of $0.0125$ motivated by the measured values of 1862 Apollo $Y = 0.022$ \citep{Kaasalainen:2007hq} and 54509 YORP $Y = 0.005$ \citep{Taylor:2007kp}. In~\citet{Rossi:2009kz}, the results were found to be invariant on the order of the uncertainty of the model to the particular distribution used. The YORP coefficient is re-drawn whenever the obliquity changes by more than $0.2$ rad and evolves according to: $Y_\text{new} =  Y_\text{old} \left( 3 \cos^2 \Delta \epsilon - 1 \right) / 2$ for smaller changes in the obliquity due to collisions as in~\citet{Nesvorny:2008by}. A similar scheme was utilized in the past~\citep{Scheeres:2007kv,Rossi:2009kz,Marzari:2011dx}. If the YORP coefficient $Y < 0$, then the spin rate is decelerating and the asteroid may enter a tumbling state. Since this model cannot assess the evolution of this state, an artificial lower spin barrier is enforced. Asteroids have a set maximum spin period limit of $10^5$ hours. At this very slow rotation rate the YORP torque switches directions. This is modeled by switching the sign of the YORP coefficient. Collisions often control the spin state of bodies with such low rotation rates since even the smallest projectiles can deliver impulsive torques that are the same order of magnitude as the angular momentum of the target body. 

The effects of collisions on the rotation rate follows a similar protocol as~\citet{Marzari:2011dx}. The population of potential impactors is derived from the Sloan Digital Sky Survey size-frequency distribution of asteroids~\citep{Ivezic:2001ct} distributed over logarithmic size bins\footnote{Diameter bins are created so that the upper diameter of a bin is $D_i = D_m D_w ^i$, where $D_m$ is the minimum diameter and $D_w = 1.25992$ is the bin width.  This is similar to \citet{Spaute:1991hv}.} from $1$ m to $40$ km. Using Poisson statistics, the number of collisions and their timing is computed for each asteroid with projectiles from each size bin using the intrinsic probability of collision for the Main Belt $\left< P_i \right> = 2.7 \times 10^{-18}$ km$^{-2}$ yr$^{-1}$ \citep{Farinella:1992im,BottkeJr:1994kr}. Each collision is assigned an impact velocity of $5.5$ km s$^{-1}$~\citep{BottkeJr:1994kr} and a random geometry within the limits of the Main Belt orbital distribution\footnote{The strongest constraint is on the velocity along the absolute z-axis which cannot exceed that predicted by the average inclination of the Main Belt.}, in order to determine from these parameters the change in spin rate due to each collision. 

Cratering collisions do not appreciably change the mass or size of the target asteroid, but they do change the angular momentum of the asteroid. The angular momentum of the projectile and the target and the geometry of the collision determine the new angular momentum of the cratered asteroid. This new angular momentum vector is used to update both the spin rate and the obliquity. Sub-catastrophic impacts create a random walk in spin rate if there is no significant YORP effect rotational acceleration~\citep{Marzari:2011dx}. If the collision is too large for a cratering event, then the original asteroid is shattered and a new object is created with the same size but a new initial spin state and YORP coefficient. Shattering collisions are defined as those that deliver specific kinetic energy greater than the critical specific energy of the target, which defined as the energy per unit target mass delivered by the collision required for catastrophic disruption (i.e. such that one-half the mass of the target body escapes)~\citep{Benz:1999cj,Davis:2002ts}.

Asteroid system destruction whether through a catastrophic collision or rotational disruption is a mass transfer from one size asteroid (the progenitor in the case of a binary) into two or more smaller size bodies. Each asteroid in the asteroid rotational evolution model resides in a logarithmic diameter bin and the model tracks this mass flow from larger bins into smaller bins after each destructive event. This mass flow from large asteroids into smaller asteroids is a well-studied phenomena in the context of collisional evolution of an asteroid population~\citep[e.g.][]{Davis:1979wc,Davis:2002ts,CampoBagatin:1994ki,Marzari:1995ga,Obrien:2003jk,BottkeJr:2005gd}. 

After a destructive event, the asteroid is replaced with another asteroid from the original diameter bin. This replacement is motivated by the constant flux of material into the original bin from even larger bins~\citep{Farinella:1992wn,Marzari:2011dx}. In this way, the asteroid rotational evolution model maintains a steady-state size-frequency distribution. Therefore it does not feature collisional evolution with full feedback, but the output of the asteroid rotational evolution model includes destruction statistics that we then incorporate into a collisional evolution model~\citep{Davis:1989vn,Davis:2002ts}. From which, we generate a new size-frequency distribution. Thus the asteroid rotational evolution model and the collisional evolution model complete a cycle. New  impact probabilities and a new projectile distribution could be generated from the size-frequency distribution output from the collisional evolution code and with these inputs the asteroid rotational evolution model could be re-run. This iterative process could be followed multiple times refining the results with each iteration. These iterations are left for future work. For now, we use the tracked mass flow from the asteroid rotational evolution model in the collisional evolution model to determine the first order corrected size-frequency distribution due to rotational fission.

\subsection{The collisional evolution model}
To evaluate the effects of YORP fissioning on the overall collisional evolution of asteroids in the Main Belt, we have used a simple 1--dimensional collisional evolution code~\citep{CampoBagatin:1994ki,Marzari:1995ga,BottkeJr:2005gd} . The size distribution of the Main Belt asteroids is modeled by a set of discrete logarithmic bins spaced by a factor 2 in mass and the time evolution is simulated through a sequence of timesteps. At each timestep the expected number of collisions involving bodies belonging to any pair of different bins is computed from  $\left< P_i \right>$. This modeled distribution is appropriate for comparison to the Main Belt after 3.8 Gyr of evolution (to account for the possible Late Heavy Bombardment~\citep{Tera:1973tf}) or after 4.5 Gyr. The outcome of each collision is computed in terms of cratering or breakup and all the fragments, at the end of the timestep, are allocated in their new size bins. The number of asteroids is then updated in each bin according to the results of the mutual collisions. The size-strength scaling adopted in the model is similar to that described in~\citet{BottkeJr:2005gd}. Our innovative approach consists in including in this model the additional erosion mechanism related to the rotational disruption. The rotational evolution code described in the previous section gives as output the frequency of rotational disruption events for different asteroid sizes. We use this frequency to compute for each size bin of the collisional evolution code the number of bodies undergoing fission during the timestep and we subtract this number from each bin adding, at the same time, the fragments to the lower size bins. Their relative sizes are chosen from a flat size distribution. Once we run the collisional evolution code with the additional grinding mechanism related to YORP, we produce a size-frequency distribution that we compare to observations. 

These smaller fragments can also be the two components of the binary created by the rotational disruption. In this case, it is possible that the rotational evolution of the binary members is significantly effected by their membership. Since $\sim$15\% of small asteroids (diameters between 0.3 and 10 km) are binaries~\citep{Pravec:1999wt,Margot:2002fe}, this effect is an important second-order effect to be dealt with in future work.

\section{Results: size-frequency distribution}
\label{sec:sizefrequencydistribtuion}

\begin{figure}
\centering
\includegraphics[width=\columnwidth]{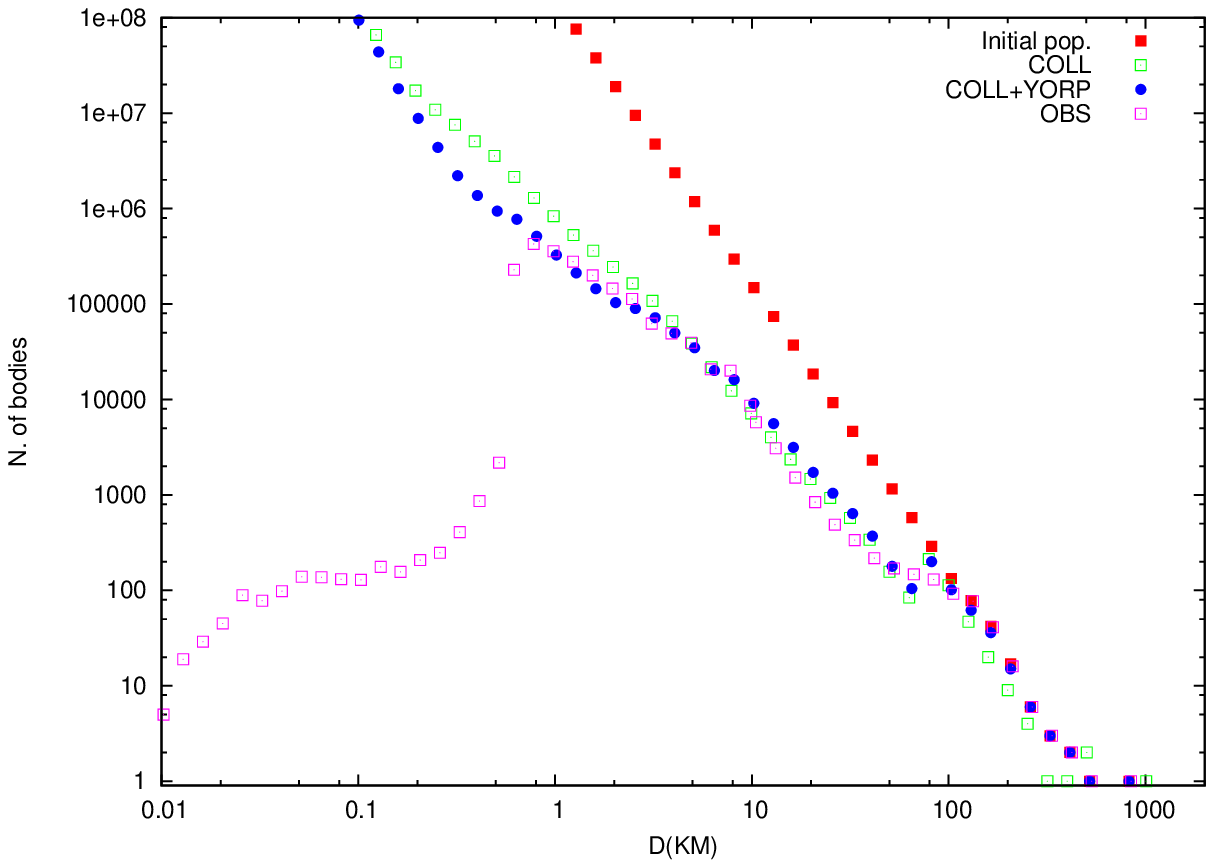} \\
\includegraphics[width=\columnwidth]{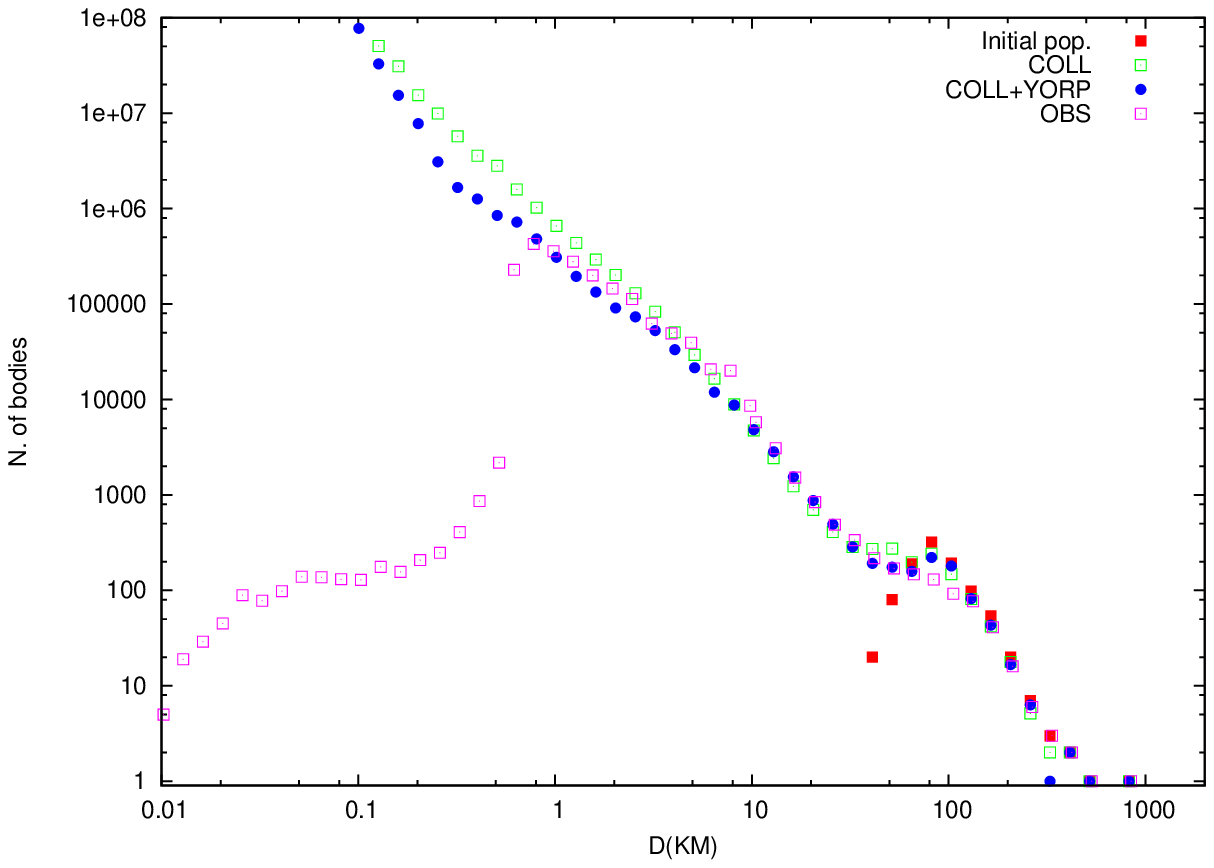}
\caption{Incremental size-frequency distributions corresponding to different initial distributions. Each is assumed for the Main Belt at the end of the accretion phase or after the Late Heavy Bombardment. They are either a steep power-law $N\left( > R \right) \propto R^{-4}$ for the top model~\citep{Weidenschilling:2011kt} or the same but smoothly truncated below $D = 100$km for the bottom model to simulate the scenario where `Asteroids were born big'~\citep{Morbidelli:2009dd}. In both plots the initial size distribution is shown as red squares. The observed distribution, reference for the modeling, is shown as magenta empty squares and it is computed by plugging in the results from the SKADS survey~\citep{Gladman:2009cx} to the~\citep{BottkeJr:2005gd} size distribution at 10 km in diameter. The size-frequency distribution of the collisional evolution model, without rotational fission, is shown on both panels as green empty squares after 3.8 Gy. The outcome of the complete model is shown as blue stars. The effects of rotational breakup begins at $D ~ 15$km in diameter but their effect is noticeable for diameters less than 6 km. Continuing both simulations to 4.5 Gy does not significantly change the results.}
\label{fig:sizefrequencyplots1}
\end{figure}
The asteroid rotation evolution model and the collision model evolve initial asteroid populations into size-frequency distributions that share many of the features of the observed size-frequency distributions by~\citet{Gladman:2009cx}. Figure~\ref{fig:sizefrequencyplots1} show the initial final size-frequency distributions for two different initial populations. The initial size-frequency distributions are shown as red squares and represent either a canonical accretion scenario (top)~\citep{Weidenschilling:2011kt} or an ``Asteroids were born big'' scenario (bottom)~\citep{Morbidelli:2009dd}. 

From each scenario, we conducted two experiments. First, asteroids did not rotationally evolve due to the YORP effect and so there was no YORP-induced rotational fission. These are the green empty squares in Figure~\ref{fig:sizefrequencyplots1}. Second, asteroids did rotationally evolve due to the YORP effect, and so there was YORP-induced rotational fission. These are the blue circles in Figure~\ref{fig:sizefrequencyplots1}. Consistent with the prediction from the simple timescale analysis, these size-frequency distributions are very similar for asteroids with radii $R \gtrsim 6$ km regardless of the initial asteroid population. Collisions solely determine the size-frequency distribution equilibrium at large sizes, where the YORP effect is irrelevant.

However, at radii $R \lesssim 6$ km, these model size-frequency distributions diverge. The second experiment, which included YORP-induced rotational fission, has far fewer asteroids in each size bin than the first experiment, which did not include YORP-induced rotational fission. This shallowing of the size distribution reflects a new equilibrium. The asteroid population at this new equilibrium experiences enhanced destruction due to rotational disruption. In other words, the collisional cascade, which produces the mass within these bins, is not able to produce new asteroids fast enough. Since this model does not include a full feedback loop, it is possible that this deficit of smaller asteroids will influence the destruction rate of asteroids that refill these size bins. This effect is likely to be small since most new mass in smaller bins is the result of catastrophic impacts between asteroids of similar sizes.

This new equilibrium matches observations. The results from the Sub-Kilometer Asteroid Diameter Survey (SKADS) survey are plotted as purple open squares and extend past the $\sim18$ magnitude ($\sim0.86$ km) limit of the survey~\citep{Gladman:2009cx}. The SKAD data transition at a radius of $\sim5$ km corresponds very close to the transitions observed in the models including YORP-induced rotational fission.

We point out that it is out of the scope of this paper to perform an accurate study of the influence on the collisional evolution of different scaling laws for the strength or of the collisional parameters. Our goal is to select a 'standard' case that reproduces reasonably well the observed size distribution at large diameters and plug in the rotational disruption algorithm to test its effect at the small size end. We intend to point out that rotational fission is an additional mechanism that must be accounted for in modeling the evolution of the Asteroid Belt.

\section{Conclusions}
Rotational disruption is a new size-dependent mechanism that alters the collisional steady-state size-frequency distribution equilibrium of Main Belt asteroids. We find that this mechanism becomes important at radii $R\lesssim6$ km from both a timescale analysis and a detailed numerical model. It nicely explains the  change to a shallower slope observed in the size distribution of asteroids at small sizes (i.e. SKAD). This finding appears to be robust since we obtain the same result even with different initial size distribution populations. Future modeling of the collisional evolution of asteroids must include this effect in their algorithm.

\section*{Acknowledgments}
SAJ would like to thank the NASA Earth and Space Science Fellowship program.

\bibliographystyle{mn2e.bst}
\bibliography{biblio.bib}

\bsp

\label{lastpage}

\end{document}